\documentclass[prl,aps,showpacs,superscriptaddress,preprint,floatfix,
nofootinbib]{revtex4}

\usepackage{amsmath,amsfonts,amssymb}
\usepackage{graphicx}

\def\3{2.8in}    
\def\2{2.5in}
\def\4{3.0in}

\def \beq {\begin{equation}}
\def \eeq {\end{equation}}
\begin{document}

\title{Discovery of several large families of Topological Insulator classes with backscattering-suppressed spin-polarized single-Dirac-cone on the surface}
\author{Su-Yang Xu}\affiliation {Joseph Henry Laboratory, Department of Physics, Princeton University, Princeton, New Jersey 08544, USA}\affiliation {Princeton Center for Complex Materials, Princeton University, Princeton, New Jersey 08544, USA}
\author{L. A. Wray}\affiliation {Joseph Henry Laboratory, Department of Physics, Princeton University, Princeton, New Jersey 08544, USA}
\author{Y. Xia}\affiliation {Joseph Henry Laboratory, Department of Physics, Princeton University, Princeton, New Jersey 08544, USA}
\author{R. Shankar}\affiliation {Joseph Henry Laboratory, Department of Physics, Princeton University, Princeton, New Jersey 08544, USA}
\author{A. Petersen}\affiliation {Joseph Henry Laboratory, Department of Physics, Princeton University, Princeton, New Jersey 08544, USA}
\author{A. Fedorov}\affiliation {Lawrence Berkeley National Laboratory, Berkeley, California 94305, USA}
\author{H. Lin}\affiliation {Department of Physics, Northeastern University, Boston, Massachusetts 02115, USA}
\author{A. Bansil}\affiliation {Department of Physics, Northeastern University, Boston, Massachusetts 02115, USA}
\author{Y. S. Hor}\affiliation {Department of Chemistry, Princeton University, Princeton, New Jersey 08544, USA}
\author{D. Grauer}\affiliation {Department of Chemistry, Princeton University, Princeton, New Jersey 08544, USA}
\author{R. J. Cava}\affiliation {Department of Chemistry, Princeton University, Princeton, New Jersey 08544, USA}
\author{M. Z. Hasan}\affiliation {Joseph Henry Laboratory, Department of Physics, Princeton University, Princeton, New Jersey 08544, USA} \affiliation {Princeton Center for Complex Materials, Princeton University, Princeton, New Jersey 08544, USA} \affiliation {Princeton Institute for Science and Technology of Advanced Materials, Princeton University, Princeton, New Jersey 08544, USA}



\maketitle

\textbf{Three dimensional topological insulators are novel states of quantum matter that feature spin-momentum locked helical Dirac fermions on their surfaces \cite{Moore Nature insight, Zahid RMP, Kane PRL, Moore PRB, David Nature BiSb, David Science BiSb, Matthew Nature physics BiSe, David PRL BiTe, Chen Science BiTe, Hor PRB BiMnTe, David Nature tunable, Hor PRB BiSe, Wray arXiv CuBiSe} and hold promise to open new vistas in spintronics, quantum computing and fundamental physics. Experimental realization of many of the predicted topological phenomena requires finding multi-variant topological band insulators which can be multiply connected to magnetic semiconductors and superconductors \cite{Hor PRB BiMnTe, Wray arXiv CuBiSe, David Nature tunable, Roy PRB, Qi PRB, Liang Fu PRL Superconductivity, Qi Science Monopole, Essin PRL Magnetic, Linder PRL Superconductivity, Franz Nature material Heusler news, Hsin Nature material Heusler, Zhang Nature material Heusler, Hor PRB BiSe, Dzero PRL Kondo}. Here we present our theoretical prediction and experimental discovery of several new topological insulator classes in AB$_2$X$_4$(124), A$_2$B$_2$X$_5$(225), MN$_4$X$_7$(147), A$_2$X$_2$X'(221) [A,B=Pb,Ge,Sb,Bi and M,N=Pb,Bi and X,X'=Chalcogen family]. We observe that these materials feature gaps up to about 0.35eV. Multi-variant nature allows for diverse surface dispersion tunability, Fermi surface spin-vortex or textured configurations and spin-dependent electronic interference signaling novel quantum transport processes on the surfaces of these materials. Our discovery also provides several new platforms to search for topological-superconductivity in these exotic materials.}

The crystal structures of AB$_2$X$_4$, A$_2$B$_2$X$_5$, and MN$_4$X$_7$ are composed of X layers forming a cubic close packing, with a fraction of octahedral interstices occupied by A and B atoms \cite{GBT Structure 1} (Fig.1). The unit cell of AB$_2$X$_4$ is formed by stacking three 7-atomic-layer-slabs in the sequence X(1)-B-X(2)-A-X(2)-B-X(1) together \cite{GBT Structure 1}. We present our first-principle theoretical calculations of the (111) surface electronic structure of AB$_2$X$_4$, A$_2$B$_2$X$_5$, MN$_4$X$_7$, A$_2$X$_2$X' respectively, along the $\bar{K}-\bar{\Gamma}-\bar{M}$ momentum-space trajectories. Results reveal a singly degenerate gapless surface state Dirac cone centered at the $\bar{\Gamma}$ point for PbSb$_2$Te$_4$, PbBi$_2$Se$_4$, GeBi$_2$Te$_4$, Pb$_2$Bi$_2$Se$_5$, PbBi$_4$Te$_7$, Sb$_2$Te$_2$Se, Bi$_2$Se$_2$S, and Sb$_2$Te$_2$S (see online Supplementary Information (SI) for the rest), indicating that these materials belong to the $Z_2=-1$ topological insulator class. It is interesting to note that crystals of MN$_4$X$_7$, although possessing bulk inversion symmetry, can feature two possible surface terminations along the (111) direction (labeled I and II in Fig. 1f). In our calculations these two variants of the PbBi$_4$Te$_7$ (111) surface possesses distinct band dispersions and anisotropy. A significantly greater charge density is expected in the top atomic layers of surface II by comparison with surface I, suggesting that the application of a Coulomb potential gradient during evaporative growth or cleavage can generate the desired surface or an alloyed combination of the two. By contrast, a fully gapped system without any surface state is observed in GeSb$_2$Te$_4$. We hence predict that GeSb$_2$Te$_4$ is topologically trivial at its experimental lattice constants. Our theoretical results show a wide range of variability in electronic and spin properties among these materials. The theoretical bulk band gap varies over an order of magnitude from $0.01eV$ to $0.31eV$ (experimental band-gaps are larger, see below). The topological surface electron kinetics range from a nearly isotropic Dirac cone (e.g. PbBi$_2$Se$_4$) to strongly anisotropic (hexagonal warping etc.) and doping-dependent electron kinematics on the PbBi$_4$Te$_7$ surface I.

We have grown several of these materials in their single crystal forms. We present experimental results on a few representative compounds (the rest will be presented elsewhere) that exhibit the clearest surface features: GeBi$_2$Te$_4$(GBT124), Bi$_2$Te$_2$Se(BTS221) and Sb$_2$Te$_2$Se(STS221) are the focus of this paper. The bulk crystal symmetry fixes a hexagonal Brillouin zone (BZ) for the cleaved (111) surface (Fig. 2a) on which $\bar{\Gamma}$ and $\bar{M}$ are the time reversal invariant momenta (TRIM) at which Dirac points or Kramers' nodes can occur. Band structure measurements using angle-resolved photoemission spectroscopy are presented by scanning over the full BZ. High resolution dispersion maps along $\bar{\Gamma}-\bar{M}$ and $\bar{\Gamma}-\bar{K}$ directions trace a clear single Dirac cone for GBT124 and BTS221 (Fig. 2b). The Dirac bands intersect at the Fermi level at $0.14\AA$, with a particle velocity of $3.6{\times}10^5m/s$ along $\bar{\Gamma}-\bar{M}$ and cross $E_F$ at $0.12\AA$ along $\bar{\Gamma}-\bar{K}$ with a much higher velocity of $5.0{\times}10^5m/s$. No other band feature is observed inside the Dirac cone, suggesting that the bulk conduction band minimum is above the chemical potential and the naturally occurring Fermi level lies inside the band-gap.

Other members of the family such as BTS221 and STS221 are also single Dirac cone topological insulators. Incident photon energy dependence (Fig. 4b) on BTS221 clearly verifies the surface origin of the ``V'' shaped Dirac band since it does not show any $k_z$ dispersion with incident photon energy. Band structure below the Dirac point, however, is found to change dramatically with $k_z$, indicating that it represents the bulk bands. Potassium surface deposition is performed in order to image the Dirac point of STS221. With 1.67 $\AA$ of K deposition, the lower Dirac cone bands are observed to cross each other (Fig. 4c). Therefore, we report STS221 as the first naturally p-type topological insulator to be experimentally confirmed. Remarkably in BTS221 we observe a Fermi momentum and velocity of $0.09\AA$ and $1.1{\times}10^6m/s$ along $\bar{\Gamma}-\bar{M}$, and $0.08\AA$ and $1.5{\times}10^6m/s$ along $\bar{\Gamma}-\bar{K}$. This is nearly a factor of three larger than the Fermi velocity in any other known topological insulator.

While the spin-textured Dirac states guarantee a non-zero surface Berry's phase, warping effects observed on the surface suggest three dimensional spin-textures \cite{Zahid Viewpoint}. Spin-texture determines the detailed nature of surface charge transport, since on the topological surface spin and quasiparticle momentum are locked in relative to one another. In general, deviations from the ideal Dirac cone shape of the surface states lead to three dimensional topological spin-textures. In order to systematically analyze the surface band structure and warping of GBT124, and how it could be tuned with bulk doping, we perform a series of high resolution ARPES measurements of the constant energy contours at different binding energies (Fig. 3a, b). The Fermi contour of GBT124 (constant energy contour at $E_B=0.01eV$, see Fig. 3a) is warped, demonstrating the hexagonal warping effect \cite{Liang Fu Warping}. When the binding energy is increased from the Fermi level, the effect of the bulk potential vanishes and the shape of the contour recovers to a circle (Fig. 3b). Lowering the binding energy further results in a Fermi surface of a single Dirac point with no other features. Hence unlike Bi$_2$Te$_3$ whose Dirac point is buried under portions of the lower surface Dirac band and the bulk valence band maximum \cite{David PRL BiTe, Chen Science BiTe}, GBT124 has an isolated Dirac point, which makes it possible to bring the system into a Dirac point transport regime \cite{David Nature tunable} not possible in Bi$_2$Te$_3$. Constant energy contours below the Dirac point are observed to consist of the lower cone, with an additional six-fold symmetric feature extending outside along all $\bar{\Gamma}-\bar{M}$ directions.

The unique spin-helical Dirac fermions of topological insulators embody a momentum-locked spin polarization that strongly affects electron dynamics, and is characterized by a non-trivial topological Berry's phase as demonstrated by us previously \cite{David Science BiSb, David Nature tunable}. The spin texture of GBT124 around the hexagonal Fermi contour in three dimensional space. Spin is found to develop a large out-of-plane $\sigma_z$ component which results from the hexagonal warping effect due to the bulk crystal potential. The in-plane spin component, on the other hand, follows the Fermi contour with \textit{left-handed chirality} and thus achieves a quantum phase of $\pi$. A full 3-D image of the topological surface Dirac band with the spin z-component suggests that the out-of-plane spin in GBT124 is found to oscillate around the constant energy contour with a period of $2\pi/3$.

Our comprehensive experimental measurements of the constant energy contours along with the spin-texture observations above will help gain key insights on the quasiparticle interference and scattering processes relevant for surface transport.
These experimental realizations of topological insulators reveal rich interplay of electronic, spin, quasiparticle interference and potential ordering instabilities on the topological surface in their native state. These ternaries also hold promise for observing superconductivity in doped topological insulators \cite{Wray arXiv CuBiSe}.


\newpage
\begin{figure*}
\includegraphics[width=17cm]{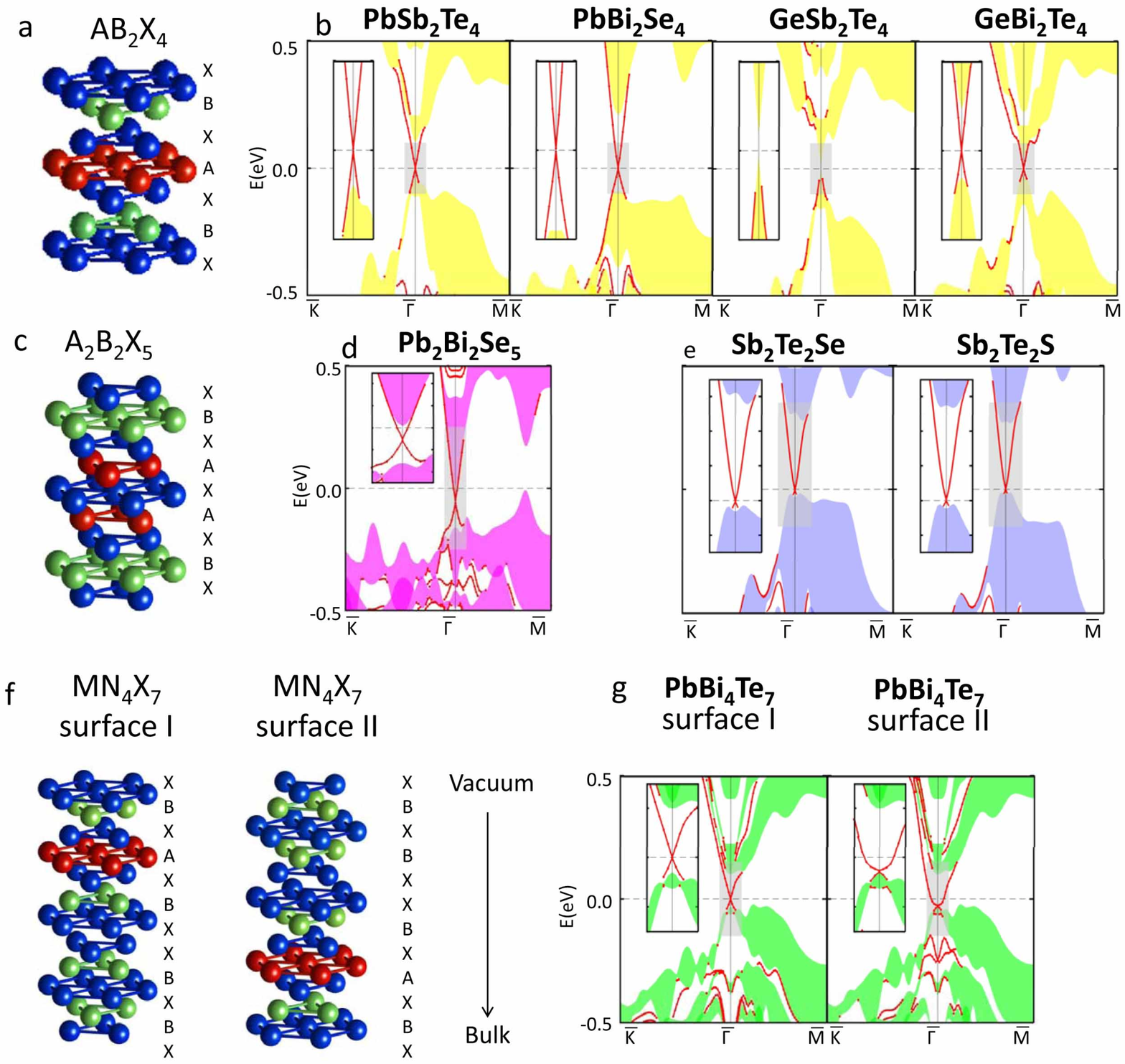}
\caption{\textbf{Topological insulator states in AB$_2$X$_4$, A$_2$B$_2$X$_5$, MN$_4$X$_7$, A$_2$X$_2$X' ternary compounds.} \textbf{a}, Crystal structure of AB$_2$X$_4$. AB$_2$X$_4$ crystal is made up of stacked 7-layer-slabs. \textbf{b}, First principle theoretical calculation of PbSb$_2$Te$_4$, PbBi$_2$Se$_4$, GeSb$_2$Te$_4$, and GeBi$_2$Te$_4$ respectively. Bulk band projections are represented by shaded areas. A single gapless surface band is observed in PbSb$_2$Te$_4$, GeBi$_2$Te$_4$ and PbBi$_2$Se$_4$. These results prove that PbSb$_2$Te$_4$, GeBi$_2$Te$_4$ and PbBi$_2$Se$_4$ belong to the $Z_2=-1$ topological class. A fully gapped system is observed in GeSb$_2$Te$_4$, which indicates that GeSb$_2$Te$_4$ is topologically trivial. \textbf{c}, Crystal structure of A$_2$B$_2$X$_5$. \textbf{d-e}, First principle calculation of Pb$_2$Bi$_2$Se$_5$, Sb$_2$Te$_2$S and Sb$_2$Te$_2$S reveals $Z_2=-1$ topological order. \textbf{f}, Crystal structure and two kinds of surface termination of MN$_4$X$_7$. \textbf{g}, Calculated bulk and surface electronic structure of PbBi$_4$Te$_7$ for each surface termination.}
\end{figure*}

\begin{figure*}
\centering
\includegraphics[width=17cm]{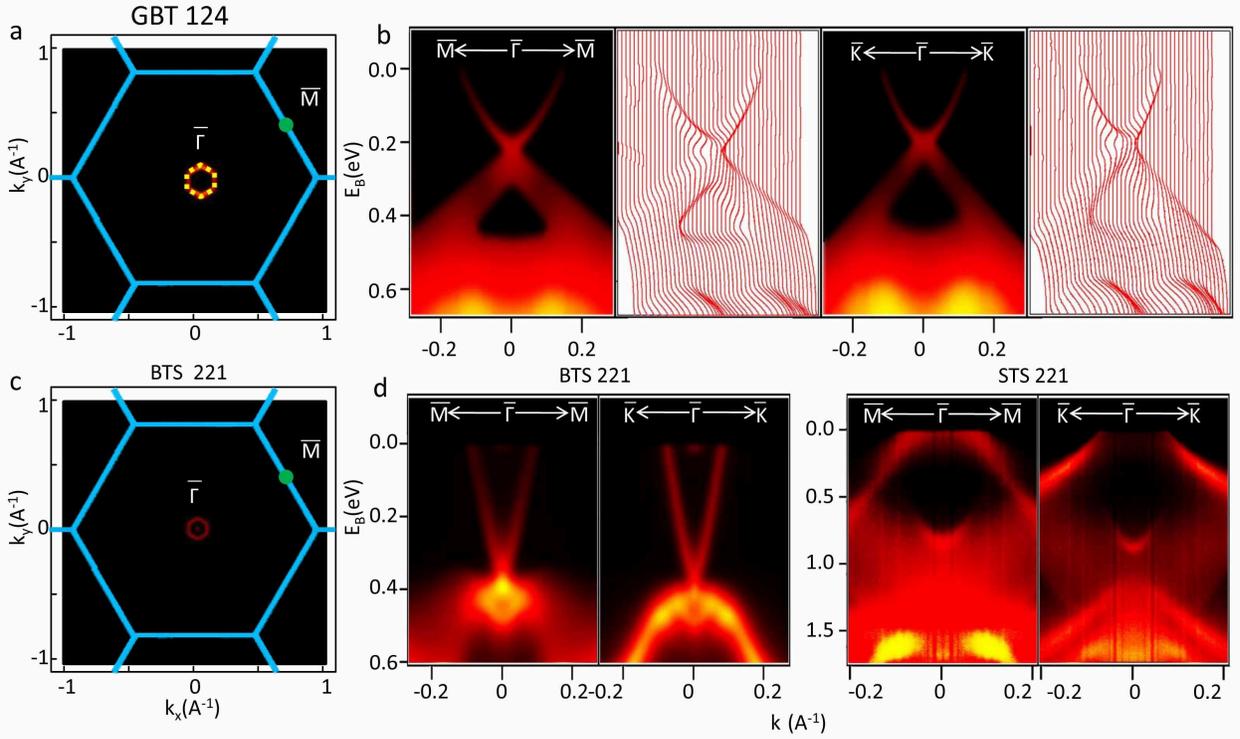}
\caption{\textbf{Long life time surface states in GBT124 and BTS221}: \textbf{a}, ARPES measurement over the First BZ on GBT124. \textbf{b}, High resolution ARPES measurements of band dispersion with the corresponding energy distribution curves along $\bar{\Gamma}-\bar{M}$ and $\bar{\Gamma}-\bar{K}$ directions. c. ARPES measurement over the First BZ on BTS221. d. High resolution ARPES measurements of band dispersion in BTS221 and STS221.}
\end{figure*}
\begin{figure*}
\includegraphics[width=17cm]{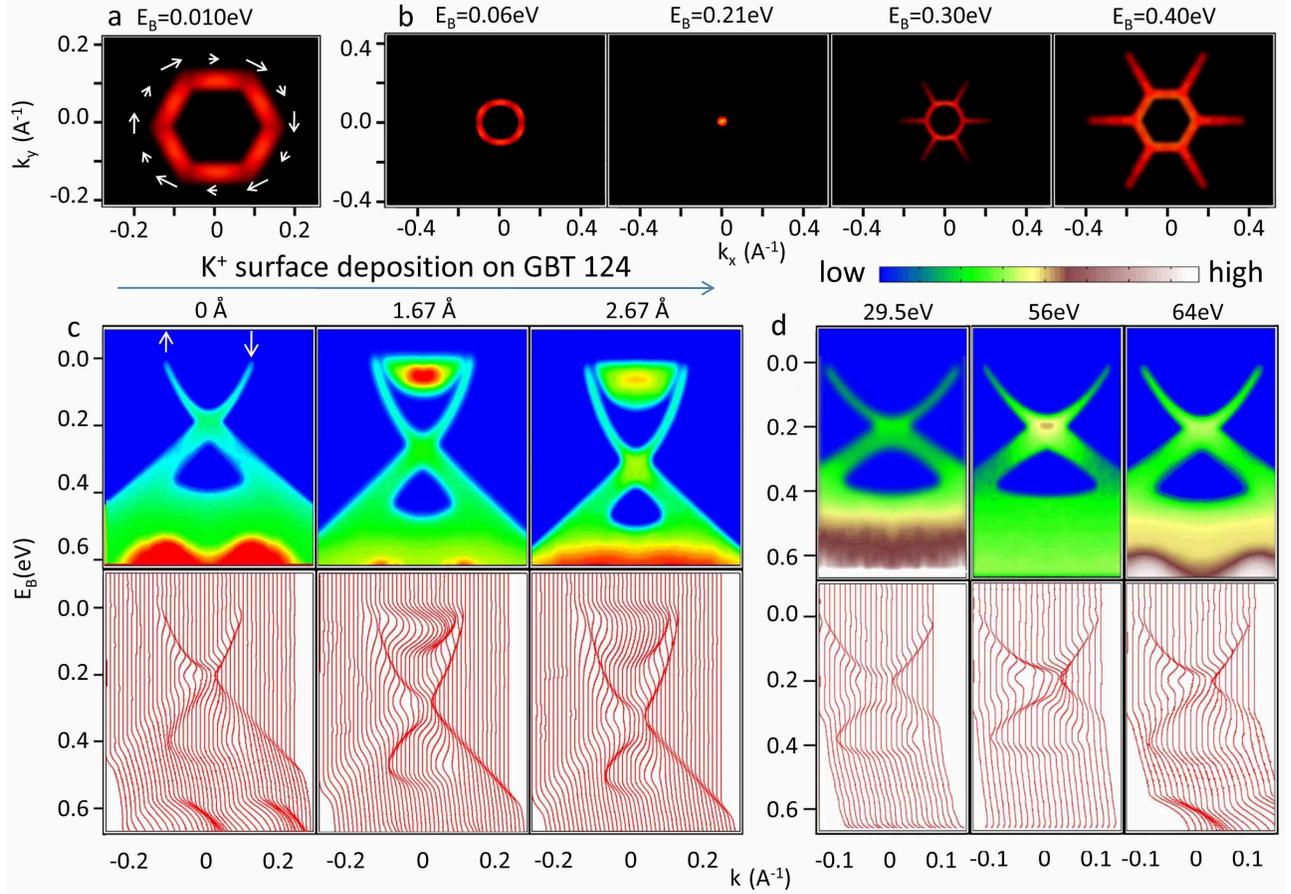}
\caption{\textbf{Evolution of topology through the Dirac node}: \textbf{a-b}, High resolution ARPES measurements of GBT124 constant energy contours at different binding energies. In-plane spin texture from the theoretical calculation is drawn on \textbf{a}. \textbf{c}, ARPES measurement of GBT124 $\bar{\Gamma}-\bar{K}$ band structure as potassium is deposited. Average thickness of the deposition layer is indicated on the top of each panel. \textbf{d}, Incident photon energy dependence of measurements along the $\bar{\Gamma}-\bar{K}$ direction. No $k_z$ dispersion is observed, which
confirms the surface character of the Dirac cone of a topological insulator.}
\end{figure*}
\begin{figure*}
\includegraphics[width=17cm]{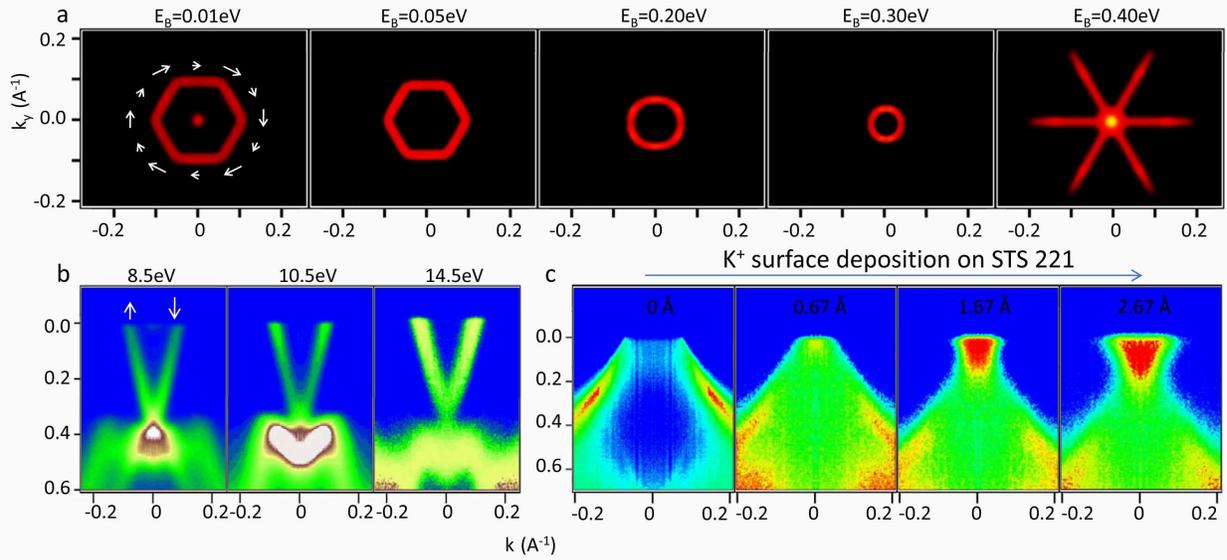}
\caption{\textbf{Warping evolution of topological states } \textbf{a}, High resolution ARPES measurements of constant energy contours on BTS221 at different binding energies. In-plane spin texture from the theoretical calculation is drawn on the first panel. \textbf{b}, Incident photon energy dependence of measurements along the $\bar{\Gamma}-\bar{M}$ direction on BTS221. No $k_z$ dispersion of the ``V'' shaped Dirac band is observed, which confirms the surface character of the Dirac cone of a topological insulator. \textbf{c}, ARPES measurement of STS221 $\bar{\Gamma}-\bar{K}$ band structure as potassium is deposited. Average thickness of the deposition layer is indicated on the top of each panel.}
\end{figure*}


\end{document}